\documentclass[%
reprint,
amsmath,amssymb,
aps,
prb,
]{revtex4-2}

\usepackage{graphicx}
\usepackage{dcolumn}
\usepackage{bm}
\usepackage[mathlines]{lineno}

\begin{document}


\title{Features and Peculiarities of Gate-Voltage Modulation of Spin-Orbit Interaction in FeCoB Nanomagnets: Insights into the Physical Origins of the VCMA Effect}


\author{Vadym Zayets}
\affiliation{National Institute of Advanced Industrial Science and Technology (AIST), Umezono 1-1-1, Tsukuba, Ibaraki, Japan}

\date{\today}

\begin{abstract}
The paper investigates the systematic dependencies of the anisotropy field and the strength of spin-orbit (SO) interaction on gate voltage in Ta/FeB/MgO nanomagnets. Our findings reveal an intriguing opposite polarity in the gate-voltage dependencies of the anisotropy field   and the coefficient  of SO interaction across all studied nanomagnets. This opposite polarity indicates that the gate-voltage modulation of spin-orbit interaction is not the primary contributor to the voltage-controlled magnetic anisotropy (VCMA) effect. Instead, the gate-voltage modulation of magnetization emerges as the most probable candidate, given its polarity aligns with the observed modulation of anisotropy. The modulation of magnetic anisotropy is influenced by two major contributions of opposite polarities, which effectively counterbalance each other and reduce the overall VCMA effect.  Optimizing the balance between these contributions could potentially lead to a substantial enhancement of the VCMA effect. Our measurements did not detect any modulation of the in-plane component of spin accumulation by the gate voltage.
\end{abstract}


\maketitle


The voltage controlled magnetic anisotropy (VCMA) effect \cite{VCMA2008MaruyamaSuzuki,VCMA2009Shiota}  refers to the phenomenon  that in a capacitor, in which one of electrodes is made of a thin ferromagnetic metal, the magnetic properties of the ferromagnetic metal changes, when a voltage is applied to the capacitor. Despite the gate voltage being applied only to the boundary of the ferromagnetic metal, it can influence the magnetic properties of the entire ferromagnet, in some cases  leading to a complete reversal of its magnetization \cite{VCMA2012ReversalSuzuki,VCMA2012ReversalOhno}. This mechanism, which allows for magnetization switching controlled by gate voltage, can serve as an effective method for data recording. When an electrical pulse reverses the magnetization direction, data is stored within the ferromagnetic metal through its two opposing magnetization states. Such a recording mechanism is fast and energy-efficient and, therefore, is promising for use in the magnetic random access memory (MRAM). 

The VCMA effect presents a fascinating yet not fully understood phenomenon in the realm of magnetism. Various plausible physical mechanisms have been proposed to explain its mechanism \cite{VCMA2008Tsymbal1stPrinc,VCMA2009Tatsuki1stPrinc}. It is currently understood that the gate voltage has the capability to influence solely the interfacial properties of the nanomagnet, leaving the bulk properties unaffected.

Within a metal, the electric field is shielded by conduction electrons, preventing its deep penetration into the material.  Consequently, the voltage applied to the dielectric gate at the nanomagnet interface can only permeate and influence the few uppermost atomic layers of the metal near the gate. Nevertheless, the alteration of magnetic properties in these uppermost layers by the gate voltage can exert a significant influence on magnetic characteristics of the whole nanomagnet.

Such a significant interface-related effect can only occur when the gate voltage influences the interfacial perpendicular magnetic anisotropy (PMA), given its considerable impact on the overall magnetic properties of the nanomagnet. However, the underlying reasons and specific details of how and why the gate voltage alters PMA remain unclear, adding to the complexity of understanding the VCMA effect. Further exploration and research are required to unravel the intricacies of this intriguing phenomenon.

Spin-orbit interaction (SO) is a fundamental phenomenon that plays a critical role in determining the existence of PMA. The substantial modulation of PMA suggests that the gate voltage modulates SO strength. However, whether this modulation solely explains the VCMA effect or if there are additional contributing factors remains uncertain. Spin-orbit interaction refers to a magnetic field $H_{SO}$ of relativistic origin \cite{LandauField}  experienced by an electron while moving within an electric field $E$:

\begin{equation}
	H_{so}= \frac{v}{c^2}  E
	\label{SO_basic}
\end{equation}

where $v$ is a component of the  electron velocity perpendicular to $E$.  

Recently, an innovative method for measuring the strength of spin-orbit interaction (SO) has been introduced \cite{Zayets2024SObasic}, offering valuable insights into this intricate fundamental phenomenon. The measurement method provides deep experimental insights into the physical processes  that influence spin-orbit interaction and consequently govern magnetic anisotropy. The modulation of spin-orbit interaction strength by  an electrical current \cite{ZayetsArch2024_SO_SOT} and the dependence of spin-orbit interaction on interface polarity \cite{ZayetsArch2024_SO_interface} were experimentally observed. The present paper provides a systematic investigation into the modulation of spin-orbit (SO) strength by a gate voltage, along with an exploration of the mechanisms through which the gate voltage impacts SO strength.

\begin{figure}[tb]
	\begin{center}
		\includegraphics[width=9 cm]{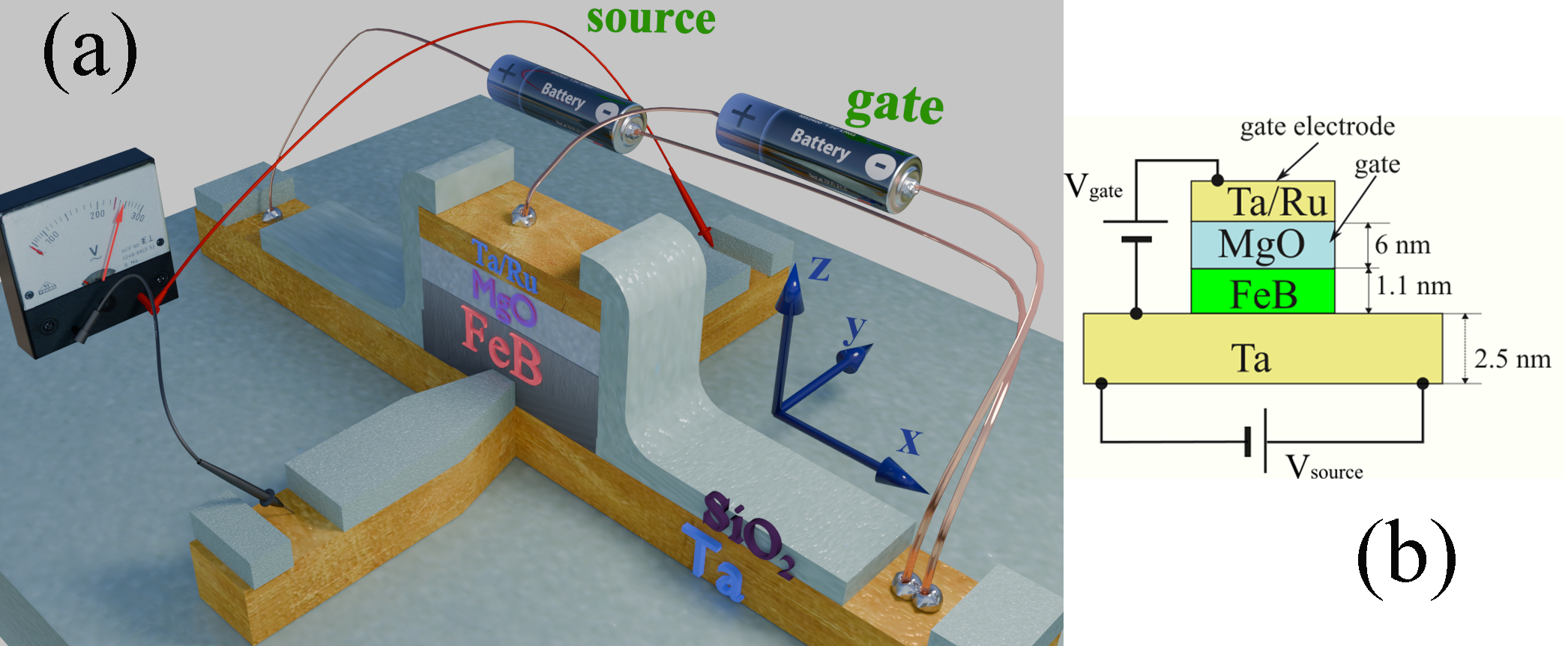}
	\end{center}
	\caption{
		(a) Experimental setup for measuring strength of spin- orbit interaction of FeB nanomagnet under gate voltage. (b)  Layer stack 
	}
	\label{fig:FigExp} 
\end{figure}


The measurement method is rooted in a fundamental characteristic of spin-orbit interaction, which is its manifestation only in the presence of broken time-reversal symmetry (T-sym), which is the case, for example,  in presence of  an external magnetic field.  This dependency of SO strength  on the degree of broken T-sym leads to an enhancement in the strength of spin-orbit interaction under an external magnetic field $H_{ext}$. Since in the absence of $H_{ext}$, T-sym remains unbroken and $H_{so}$ equals zero, there exists a linear relationship  between $H_{so}$ and $H_{ext}$:

\begin{equation}
	H_{so}=k_{so} \cdot H_{ext}
	\label{HsoVsH}
\end{equation}
where $k_{so}$ is the coefficient of spin- orbit interaction, which defines the strength of spin-orbit interaction.

Since the strength of the magnetic anisotropy is inherently linked to the strength of the spin-orbit interaction, the most intuitive and direct approach for a measurement of the SO strength is through measurements of magnetic anisotropy. Indeed, both theoretical and experimental evidence \cite{Zayets2024SObasic} has shown that the anisotropy field $H_{ani}$ linearly increases with an increase in the external magnetic field $H_z$ applied along the magnetic easy axis. This relationship is found to be as:

\begin{equation}
	H_{ani}-H_z= H^0_{ani}+k_{so}H_z
	\label{EqHani}
\end{equation} 

where $H^0_{ani}$ is the anisotropy field in absence of $H_z$. A fitting of experimental measurements of $H_{ani}$ versus $H_z$ by linear Eq. \ref{EqHani} yields values for $k_{so}$ and $H^0_{ani}$.

\begin{figure}[tb]
	\begin{center}
		\includegraphics[width=7 cm]{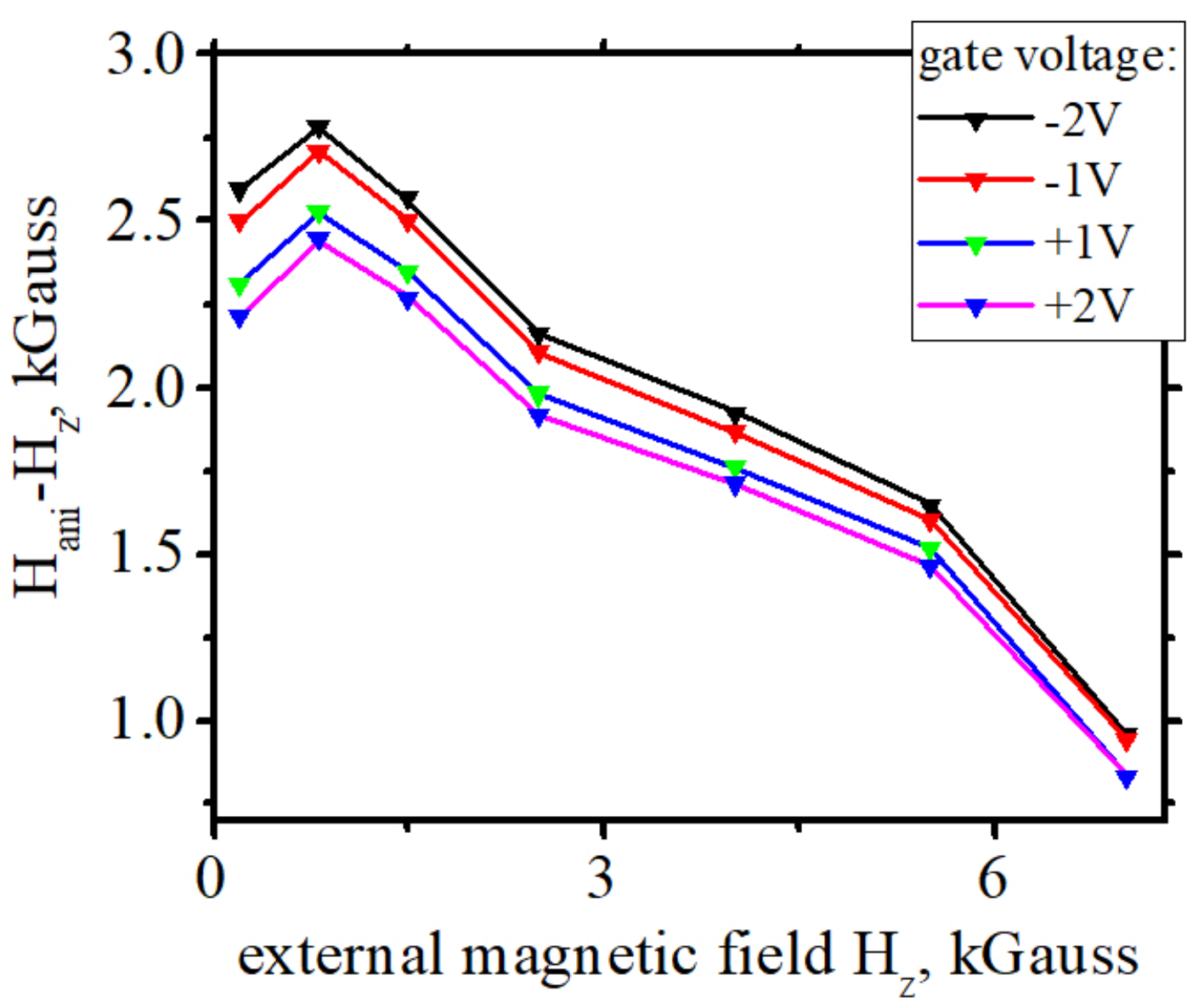}
	\end{center}
	\caption{
		{Anisotropy field $ H_{ani} $ as a function of external perpendicular-to-plane magnetic field $ H_{z} $ measured at a different gate voltage $ V_{gate} $.} 
	}
	\label{fig:FigHaniGate} 
\end{figure}


In order to investigate the dependency of the spin-orbit (SO) strength on gate voltage, a 1.1-nanometer-thick FeB nanomagnet was fabricated atop a 400-nanometer-wide Ta nanowire. A Ta- made Hall probe was aligned with the nanomagnet. Additionally, a 6-nanometer-thick MgO gate oxide layer and Ta gate electrode were constructed on top of the nanomagnet. A bias voltage is applied across the two ends of the nanowire, while a gate voltage is applied between the top gate electrode and one end of the nanowire (See Fig. \ref{fig:FigExp}). Nanomagnets are specifically utilized due to their lack of magnetic domains, simplifying data analysis. However, this measurement method allows for its application to larger samples without significant issues.

The experiments were carried out at room temperature, well below the Curie temperature of FeB. The magnetization angle was measured by a Hall probe (See Fig. \ref{fig:FigExp}).The in-plane and perpendicular- to- plane components of the applied magnetic field  are controlled individually. The measurement process involved recording the Hall angle $\alpha_{Hall}$ during the scanning of an in-plane external magnetic field $H_x$ in two opposite directions. The perpendicular-to-plane magnetic field was employed as a parameter. 

The anisotropy field $H_{ani}$ was evaluated by fitting the linear relationship \cite{Zayets2024SObasic,Johnson1996Hani} between the in-plane magnetization component $M_x$ and $H_x$. Even without an external magnetic field, there exists an in-plane magnetic field $H_{||}$,  causing  the magnetization to tilt from the magnetic easy axis. Both the magnetic field generated by spin accumulation and the Oersted field created by the current contribute to $H_{||}$ \cite{ZayetsJMMM2023Parametric}. To prevent any systematic errors, $H_{||}$ was carefully evaluated and factored into the fitting process. Detailed information regarding the measurement procedures can be found in Refs. \cite{ZayetsJMMM2023Parametric, Zayets2024SObasic}.

Figure \ref{fig:FigHaniGate} displays the measured relationship between $H_{ani}-H_z$ and the external magnetic field $H_z$ measured at a different gate voltage $ V_{gate} $.  The relationship shows an approximately linear trend with minor oscillations superimposed on it. The oscillations are a recognized characteristic of the spin-orbit (SO) interaction in the interfacial layer \cite{Zayets2024SObasic}.  Notably, both the slope and the offset of this linear dependence significantly vary with the gate voltage.  As  $V_{gate}$  increases, the offset decreases while the slope increases. This opposing relationship is clearly observed as a narrowing gap between the lines at higher  $H_z$

\begin{figure}[tb]
	\begin{center}
		\includegraphics[width=7 cm]{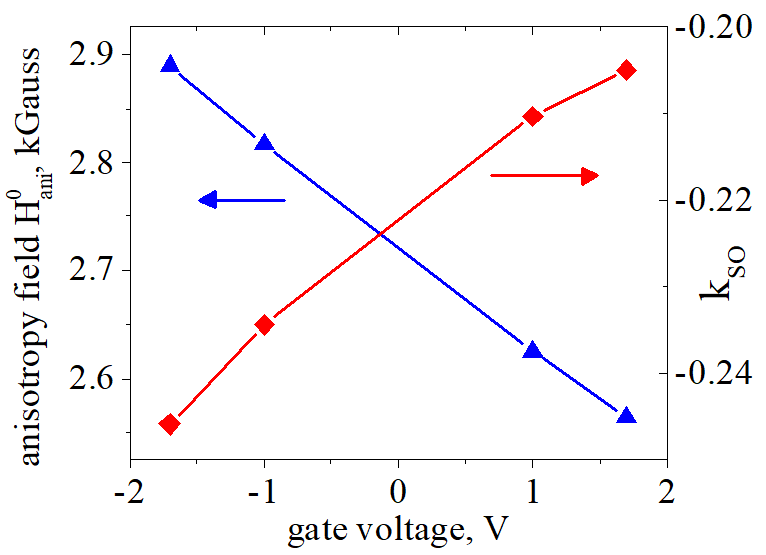}
	\end{center}
	\caption{
	Anisotropy field $H^0_{ani}$ and coefficient of the spin orbit interaction $k_{so}$ as a function of the gate voltage
	}
	\label{fig:FigKsoGate} 
\end{figure}


As shown in Eq. \ref{EqHani}, the slope is directly proportional to the spin-orbit coefficient $k_{so}$,  while the offset is proportional to the anisotropy field $H^0_{ani}$ in the absence of an external magnetic field. Figure \ref{fig:FigKsoGate} illustrates the relationship between $H^0_{ani}$ and $k_{so}$ with respect to $V_{gate}$, showcasing nearly linear dependencies for both parameters.

\begin{figure}[tb]
	\begin{center}
		\includegraphics[width=7 cm]{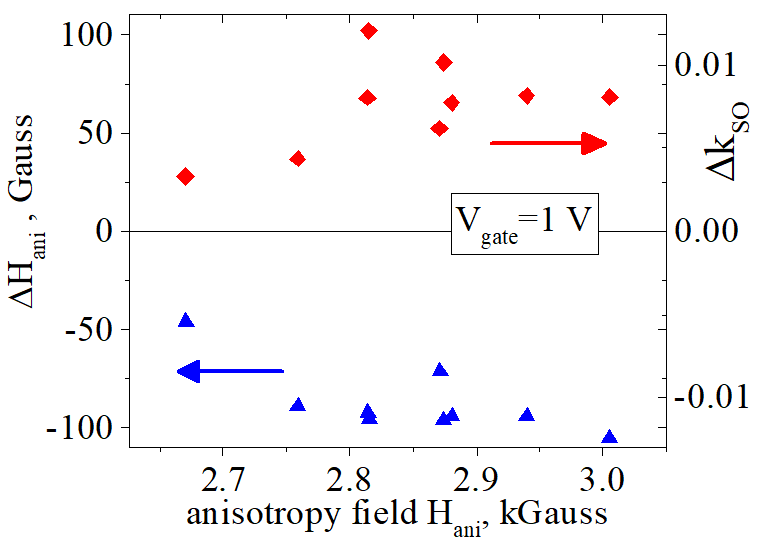}
	\end{center}
	\caption{
		The change of anisotropy field $\Delta H_{ani}$ versus the change of coefficient of SO interaction $\Delta k_{so}$ measured when gate voltage of 1 V is applied. Each dot corresponds to a measurement of one individual nanomagnets fabricated at a different place on the same wafer.
	}
	\label{fig:FigDistr} 
\end{figure}

The opposite dependencies of $H^0_{ani}$ and $k_{so}$ on $V_{gate}$ are systematic. Figure \ref{fig:FigDistr} illustrates the variations in $H^0_{ani}$ and $k_{so}$  when a gate voltage of 1 V is applied, as measured across different nanomagnets located at various positions on the same wafer. Due to variations in interface roughness, the measured values of $H^0_{ani}$ and $k_{so}$ slightly differ for each nanomagnet. In each case, $\Delta k_{so}$ is positive, while $\Delta H_{ani}$ is negative.

Additionally, similar measurements were conducted on Ta/FeB/MgO,   W/FeB/MgO, Ta/FeCoB/MgO, $[W/FeB]_n/MgO$, $[Ta/FeB]_n/MgO$ nanomagnets of a  different structure, a different material composition and a different size \cite{Intermag2023_SO_VCMA,MMM2022_SO_VCMA}. Remarkably, the observed trends remained identical. Specifically, for positive $V_{gate}$, $\Delta k_{so}$ is consistently positive, whereas $\Delta H_{ani}$ is consistently negative.

The unexpected opposite dependencies of  $H^0_{ani}$ and $k_{so}$  on gate voltage are intriguing, given that $H^0_{ani}$ and $k_{so}$  are not independent parameters. Specifically, $H^0_{ani}$  is determined by $k_{so}$  and the internal magnetic field $H_{int}$ within the nanomagnet. This internal magnetic field is responsible for maintaining magnetization along the magnetic easy axis. Given the general similarity between external $H_z$ and internal $H_{int}$ magnetic fields, one would expect their effects on the nanomagnet to be identical. Utilizing this symmetry principle, Eq. \ref{EqHani} can be rewritten as

\begin{equation}
	H_{ani}=(H_z+H_{int})+k_{so} (H_z+H_{int})
	\label{EqHaniHint}
\end{equation} 

where

\begin{equation}
	H^0_{ani}= (1+k_{so})H_{int}
	\label{EqHaniHintkso}
\end{equation} 

As shown in Eq. \ref{EqHaniHintkso},  $H^0_{ani}$  is linearly proportional to $k_{so}$, indicating that their dependencies should share the same polarity.  However, an exception arises when another parameter, in addition to $k_{so}$,  influences $H^0_{ani}$ with a gate-voltage dependency opposite to that of $k_{so}$. The magnetization is likely such a parameter, as it affects the anisotropy field and is modulated by the gate voltage. The gate voltage modulates magnetization by altering the Fermi level at the interface, which changes the filling ratio between spin-up and spin-down localized d-electrons.

The gate voltage creates  the accumulation or depletion of electrons near the gate, which leads to the modulation of the Fermi level. A positive gate voltage induces electron accumulation, raising the Fermi level. This higher Fermi level results in not only an increased number of conduction electrons but also a higher number of localized d-electrons. Given that d-orbitals in iron are more than half-filled, only unoccupied d-states with spins opposite to the total spin of iron remain. Consequently, the addition of d-electrons reduces the total spin of iron. Thus, a positive gate voltage reduces the magnetization near the gate. Conversely, a negative gate voltage induces electron depletion, thereby enhancing magnetization.

\begin{figure}[tb]
	\begin{center}
		\includegraphics[width=6.5 cm]{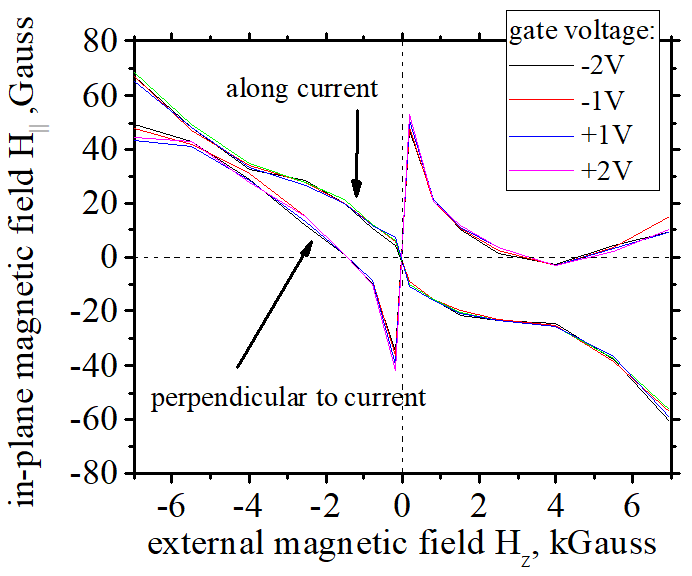}
	\end{center}
	\caption{
		The parallel-to-current and perpendicular-to-current components of the in-plane magnetic field $H_{||}$ measured under varying gate voltages.The data shows no discernible dependence of $H_{||}$ on the gate voltage within the measured precision.
	}
	\label{fig:FigHinplane} 
\end{figure}


Since the anisotropy field is linearly proportional to the magnetization \cite{Zayets2024SObasic}, a similar trend can be expected for the anisotropy field. Following the change in magnetization, the anisotropy field should decrease with an increased gate voltage. This is precisely the trend observed experimentally, as shown in Figure \ref{fig:FigKsoGate}. The experimental observation that the polarity of gate-voltage modulation of $H^0_{ani}$ coincides with the polarity of magnetization modulation, and that this contribution overrides the opposite contribution due to the modulation of spin-orbit interaction, strongly suggests that the voltage modulation of magnetization is the primary contributor to the VCMA effect.

The mechanism by which the gate voltage modulates the strength of spin-orbit interaction can be elucidated as follows. The spin-orbit interaction experienced by localized d-electrons originates from the electric field of the nucleus due to electron orbital motion. Since the d- electrons are quenched and thus  lack an orbital moment, the SO magnetic field $H_{so}$ generated by the clockwise- and counterclockwise- rotating components of their wavefunction cancels each other out, resulting in no net spin-orbit interaction. However, this balance is disrupted when internal $H_{int}$ or external $H_z$ magnetic field breaks it, leading to the emergence of overall $H_{so}$. The disparity between these two contributions to $H_{so}$, and consequently the overall $H_{so}$, substantially depends on specifics of the orbital distribution. Notably, a significant difference arises when the orbital center position diverges from the nucleus position. This explains why the spin-orbit interaction is markedly more pronounced for Fe orbitals at the interface compared to those in the bulk, a distinction clearly demonstrated in experimental measurements \cite{Zayets2024SObasic}. When a gate voltage is applied to the interface, it induces modifications in the orbital distribution. This, in turn, alters the discrepancy between the two $H_{so}$ contributions and thereby affects the overall strength of the spin-orbit interaction resulting in the gate- voltage modulation of $k_{so}$.

In the same measurement, we confirmed that the gate voltage does not modulate the amount of in-plane spin accumulation. As explained above, the in-plane magnetic field $H_{||}$ is measured simultaneously with $H_{ani}$ and $k_{so}$. $H_{||}$ serves as an indicator of the direction and magnitude of spin accumulation because the magnetic field induced by spin accumulation significantly contributes to $H_{||}$ and because the measured dynamics of $H_{||}$ closely track the dynamics of spin accumulation \cite{ZayetsJMMM2023Parametric}. Figure \ref{fig:FigHinplane} illustrates the measured parallel- to- current and perpendicular- to- current components of $H_{||}$ under varying gate voltages. No discernible dependence of $H_{||}$ on $V_{gate}$ was detected, contrasting with the substantial dependence of $H_{||}$ on the electrical current as reported in ref. \cite{ZayetsJMMM2023Parametric}. 

It's worth noting that we were only able to measure the in-plane component of spin accumulation, while the perpendicular- to- plane component of spin accumulation couldn't be assessed with used setup. However, this component can be evaluated using the measurements described in Refs. \cite{Zayets2020MishenkoSpinPol,ZayetsArch2024_SO_SOT}.

In conclusion, the systematic experimental study of spin-orbit strength in FeB nanomagnets has revealed an intriguing opposite polarity in the gate-voltage dependence of the strength of spin-orbit interaction and anisotropy field. This opposite polarity suggests that the gate-voltage modulation of spin-orbit interaction is not the primary contributor to the VCMA effect. Instead, there is a stronger contributor to the magnetic anisotropy with the opposite polarity. The gate-voltage modulation of magnetization is the most probable candidate, as its only possible polarity coincides with the experimentally observed modulation of anisotropy. Since the gate modulation of magnetic anisotropy is influenced by two major contributions of opposite polarities, they effectively counterbalance each other, reducing the VCMA effect. Optimizing the balance between these contributions could potentially lead to a substantial enhancement of the VCMA effect.


\bibliography{VCMASpinOrbitBib}

\end{document}